\documentclass{PoS}

\newcommand{\CTA}{CTA}
\newcommand{\gammacat}{\emph{gamma-cat}}
\newcommand{\hess}{H.E.S.S.}

\title{Observing the Galactic Plane with the Cherenkov Telescope Array}

\ShortTitle{GPS with CTA}

\author{CTA consortium represented by \speaker{Roberta
    Zanin}\footnote{Max-Planck Institut ur Kernphysik, Heidelberg
    (Germany)} and 
  Jamie Holder\footnote{University of Delaware, US}\\
  E-mail: \email{robertazanin@gmail.com}}

\abstract{
The Cherenkov Telescope Array is a next generation ground-based
gamma-ray observatory designed to detect photons in the 20\,GeV to
300\,TeV energy range.  With a sensitivity improvement of up to one order of
magnitude on the entire energy range with respect to currently operating facilities, coupled with
significantly better angular resolution, the array will be used to
address many open questions in high-energy astrophysics. In addition,
CTA will explore the ultra-high energy (E $>$50\,TeV) window with great
sensitivity for the first time.

CTA is expected to reveal a detailed picture of the Galactic plane at
the highest energies, and to discover around one hundred
new supernova remnants and many hundreds of pulsar wind nebulae,
according to current population estimates. The ability of the
observatory to resolve such a large number of Galactic sources is one
of the challenges to be faced. In this paper, we will present the first
simulated scan of the Galactic plane with a realistic observation
strategy, with particular attention to the potential source
confusion. We will also present prospects for morphological studies of extended
sources, such as the young SNR RX\,J1713.7-39.}

\FullConference{35th International Cosmic Ray Conference ICRC2017\\
		10-20 July, 2017\\
		Bexco, Busan, Korea}

\begin{document}

The Cherenkov Telescope Array (\CTA) is a next-generation imaging
atmospheric Cherenkov telescope (IACT) array that has been designed to
detect photons from 20\,GeV up to 300\,TeV. Currently under
construction, \CTA{} consists of two distinct arrays which will
provide complete coverage of the entire sky. CTA-North will be located on the
Canary island of La Palma, while CTA-South will be at the Paranal site
in Chile. \CTA{} will consist of multiple telescopes of three
different sizes, each of which is optimized for a specific energy
range. The largest telescopes (large-sized telescopes, LSTs), with a reflector surface of
23\,m diameter, will provide an energy threshold of few tens of GeV,
whereas the smallest (4\,m diameter small-sized telescopes, SST) 
guarantee the coverage of
the highest energies up to hundreds of TeV. Medium-sized telescopes
(12\,m diameter, MST) provide unprecedented sensitivity in the
central energy range.  The performance of the two arrays is described
in detail in \cite{Konrad,Gernot}.

\section{The CTA Galactic Plane Survey}
One of the most important legacies of \CTA{} will undoubtedly be its
Galactic plane survey (GPS). This aims to provide the first complete
very-high-energy (VHE) scan of the entire Galactic plane with a sensitivity better
than 4\,mCrab for unresolved sources. The GPS will provide a complete
census of the VHE gamma-ray sources in our Galaxy, as well as a
precise characterization of the diffuse emission in the 0.1-100\,TeV
energy range from the entire Galactic plane. Given the significant
improvement in the performance of \CTA{} with respect to current
IACTs, this dataset will likely also reveal many new and unexpected
phenomena, such as new classes of gamma-ray emitters and new types of
transient and variable source behaviour.
%and will allow to identify interesting targets for follow-up
%observations. 

\begin{figure}
\begin{center}
\includegraphics[scale=0.4]{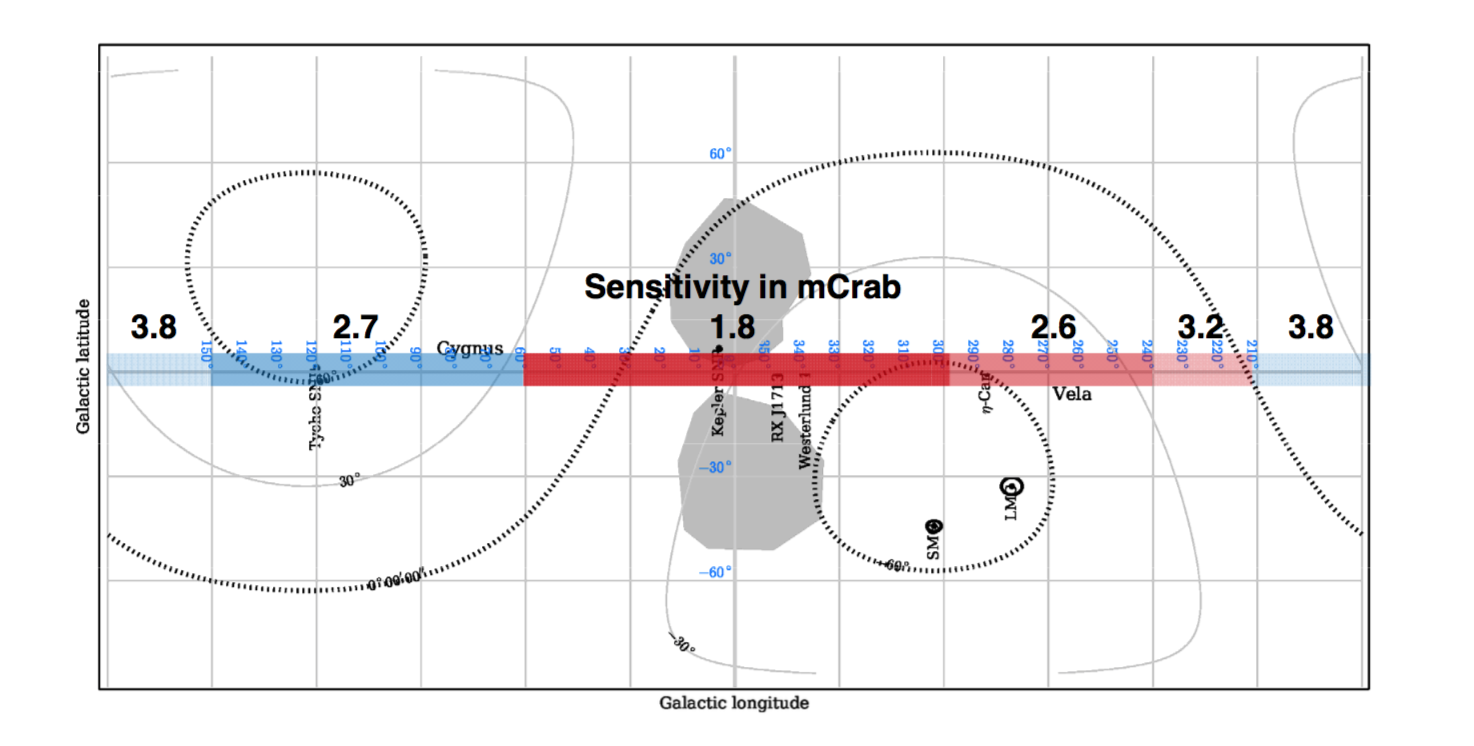}
\caption{Point-source sensitivity in mCrab foreseen for the 10-year
  GPS program along the Galactic plane. Some known VHE objects are
  labelled for reference. Figure taken from \cite{CTAScience}.}
\label{fig:sensitivity}
\end{center}
\end{figure}

The survey will be carried out using both the northern and the
southern arrays, and will provide complete coverage of the Galactic
plane. The exposure time will be non-uniform across the plane, and
will be determined by the scientific goals to be addressed in
different regions. The inner Galactic region
($60^{\circ}<l<60^{\circ}$), for instance, will be allocated
significantly more observation time than other regions, allowing to
reach a sensitivity as low as 1.8 mCrab. %This is the region....  The
GPS is divided into an early-phase (1--2\,years) and a long
term-program (3--10\,years) with a total of 1020 and 600\,hr requested
for CTA-South and CTA-North, respectively. A double-row pointing
strategy will be used, with a nominal separation distance of
3$^{\circ}$ and the observational pattern will be set to optimize
sensitivity to periodic signals on different time
scales. Figure\,\ref{fig:sensitivity} shows the different point-like
sensitivities which will be reached by the full 10-year program in the
various regions of the sky.

The current generation of ground-based instruments has revealed that more
than half of the detected TeV sources cluster along the Galactic plane
(90\% of the Galactic sources lie at $|$b$|<2^{\circ}$). Pulsar wind nebulae are the most
numerous class, followed by supernova remnants \cite{HGPS}.
However, half of the Galactic plane sources remain unidentified for three
main reasons: 1) they have multiple associations at lower energies
which cannot be disentangled, 2) they consist of several still
unresolved sources, 3) they are completely ``dark'', with no counterpart at
any other wavelength.  Counting both firm identifications and
candidate associations,  we find that 60\% of the Galactic plane sources are
PWNe-like\footnote{In this definition PWNe include composite SNRs},
and 20-25\% are SNR-like. 

\begin{figure}
\begin{center}
\includegraphics[scale=0.7]{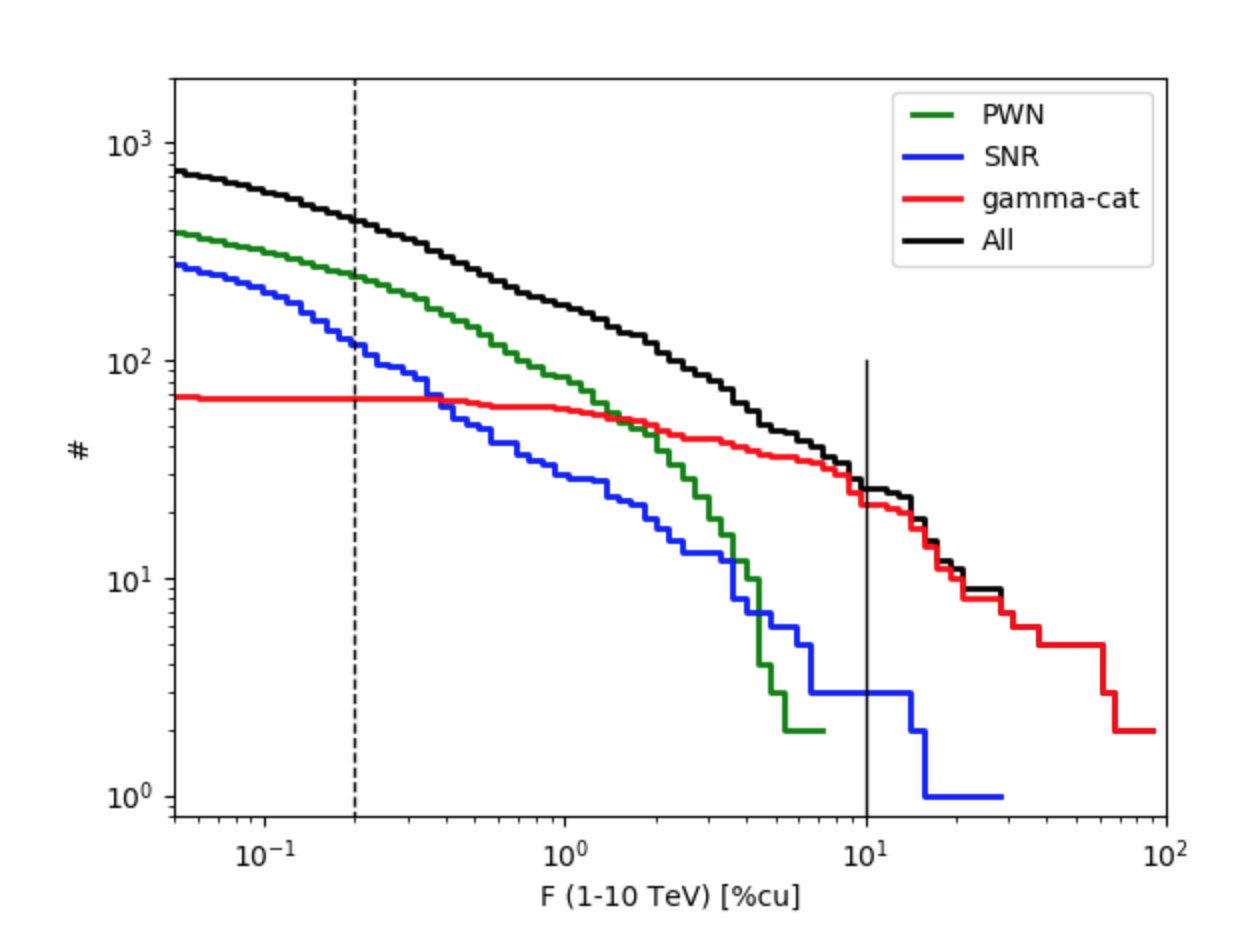}
\caption{log$N$-log$S$ distribution. The red and black lines shows the known VHE
  sources taken from \cite{gammacat} and the VHE sources expected to be
  detected by \CTA. Green and blue show the PWN and SNR populations,
  respectively, with the number of already known sources subtracted. 
The black vertical line indicates the completeness of the \hess{} GPS
at 10\% of the Crab Nebula flux, whereas the dashed line shows the
\CTA{} expected completeness, at 0.2\% of the Crab Nebula flux. }
\label{fig:logNlogS}
\end{center}
\end{figure}

Through the log$N$-log$S$ distribution of the VHE known sources, taken
from the open-source \gammacat{} catalogue \cite{gammacat}, we can
estimate the number of VHE sources expected to be detected by
\CTA. Figure \ref{fig:logNlogS} illustrates the
obtained log$N$-log$S$,  i.e. the cumulative source count as a
function of the integral flux between 1 and 10\,TeV. This predicts
less than 100 SNRs and $\sim$400 PWNe. These numbers provide only an
estimate of the number of sources expected to be detected
by \CTA{} for two main reasons: 1) the models consider point-like
sensitivities whereas the sensitivity worsens for extended sources
(as shown in \cite{Gernot}) 2) they are based on an extrapolation of the number of sources
detected by the current generation of IACTs. 

\begin{figure}
\begin{center}
\includegraphics[scale=0.5]{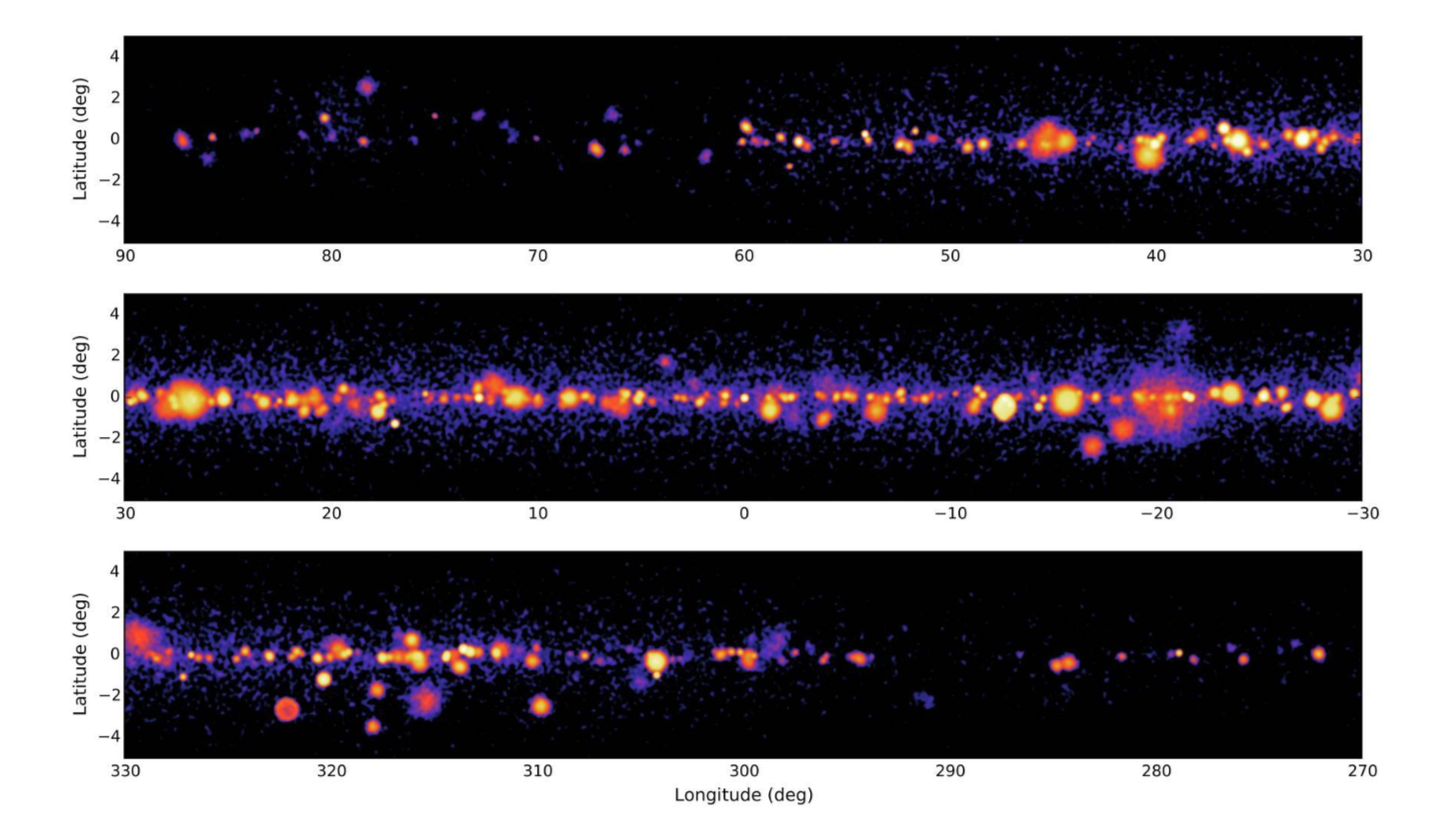}
\caption{Simulated image of the Galactic plane survey carried out from
the southern array within the 10-year program. These maps
  have not been produced by using the skymodel described in this proceeding.}
\label{fig:GPS}
\end{center}
\end{figure}

\begin{figure}
\begin{center}
\includegraphics[scale=0.5]{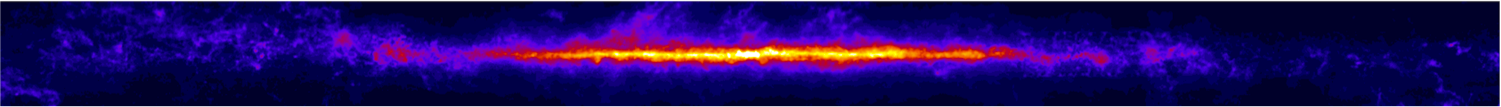}
\caption{The Galactic interstellar medium diffuse model including both
  emission from the interstellar gas and the inverse Compton
  component.}
\label{fig:diffuse}
\end{center}
\end{figure}

Figure \ref{fig:GPS} shows a realistic simulated image of the GPS obtained from
observations with the southern array. This simulation includes all of the
known VHE sources together with models
for both the PWN and SNR populations, and for the diffuse
emission. In order to preserve the total number of PWNe and SNRs, 
in agreement with the estimations derived above through the log$N$-log$S$
distribution, for each known source we removed from the
corresponding model one source with a comparable integral flux. 
The SNR model, taken from \cite{Cristofari}, accounts only for
young SNR - no interaction with the interstellar medium is foreseen
at this stage. The PWN model was built with a phenomenological
approach, similar to that described in \cite{Renaud}. In
particular, we considered the radial distribution of the surface density
of pulsars in the Galaxy taken from \cite{Yusifov}. The
intrinsic size was randomly generated by using a Gaussian probability
density function, truncated at 0. The differential energy spectra were
assumed to be log-parabola functions, with the index and curvature 
parameters were randomly generated
by using Gaussian probability density functions derived from the corresponding 
distributions of the known PWNe \cite{Kargalsev,MayerThesis}, whereas 
the luminosity is estimated from a power-law density function with a
slope of about one.  
We also coupled one-third of the PWN with the SNR to create a sample
of composite SNRs with a variable ratio between the radius of the two
components. \\
The Galactic interstellar medium diffuse model is based on predictions
from codes that solve the cosmic ray (CR) transport equations and
calculate the related multiwavelength emission. It includes emission
from interstellar gas, produced by hadronic interactions of cosmic ray
nuclei and by electron/positron Bremsstrahlung. It also accounts for
the inverse-Compton (IC) radiation produced by the interactions of
cosmic ray electrons and positrons with low-energy photons. We use
predictions from \emph{Dragon} \cite{Dragon}, that assume
position-dependent diffusion and convection properties to reproduce
the intensity and spectral hardening of interstellar gamma-ray emission seen by
\emph{Fermi}-LAT toward the inner Galaxy. The resolution of the gas
maps has been enhanced with respect to those used internally in
\emph{Dragon} by using the dust maps from Planck and IRAS, which have an
angular resolution of 0.03$^{\circ}$. The IC component was obtained
using predictions from \emph{Picard}, that assume a model with 4 spiral arms
for the CR source distribution \cite{Picard}. Figure \ref{fig:diffuse}
shows the count map of the obtained diffuse model.
% which has large imprint on the resulting IC emission  
%Besides these po, we added on top of the 6 known gamma-ray binaries, 5 new 
%weak point-like sources arbitrarily distributed along the Galactic plane 
%with either a long orbital period or a short duty cycle to justify 
%their undetectability by the currently running facilities. 

This simulation of the GPS shows that source confusion is the most limiting
factor in the detection of sources, especially in the innermost
regions of the Galaxy where the source density can approach 3-4 sources per
square degree. Preliminary studies, not yet considering the diffuse
emission, have set lower limits to the estimated confusion from
13-24\% at 100\,GeV to 9-18\% at 1\,TeV for the region $|l|<30^{\circ}$
\cite{CTAScience}.  

\section{Morphological Capabilities}

\begin{figure}
\begin{center}
\includegraphics[scale=0.45]{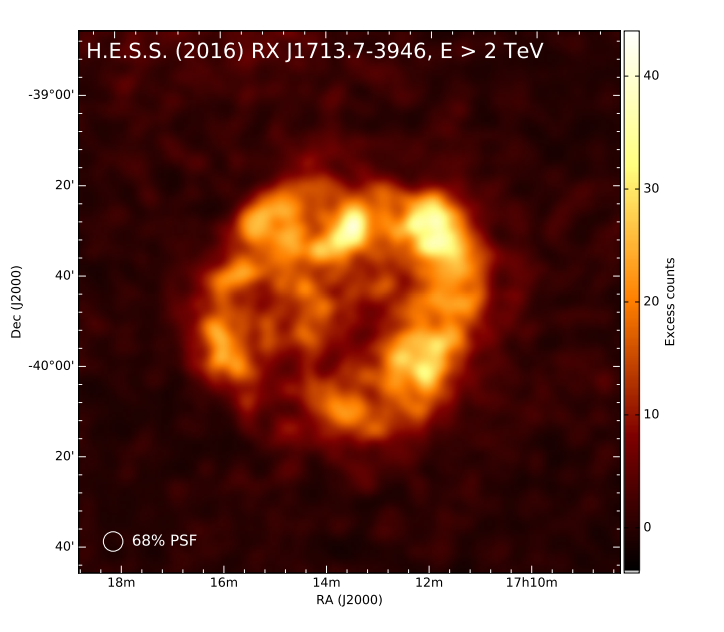}
\includegraphics[scale=0.6]{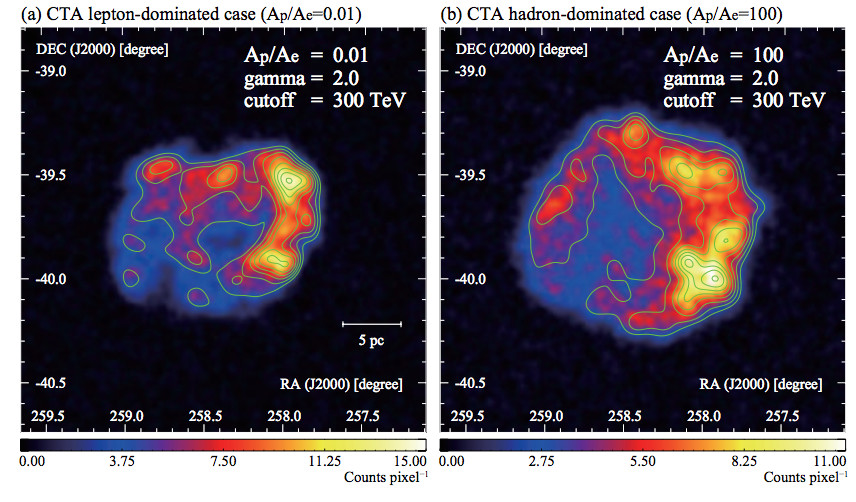}
\caption{Image of RX\,J1713.7-3946. \emph{On the right} \hess{} result
  above 2\,TeV taken from \cite{j1713}; \emph{On the left} expected 
\CTA{} results for the leptonic and hadronic dominated gamma-ray
emission \cite{j1713cta}.} 
\label{fig:rx}
\end{center}
\end{figure}

The majority of the known VHE gamma-ray sources in the Galaxy are
spatially extended, with no clear sub-structures and a typical angular
size of 0.1-0.2$^{\circ}$ \cite{HGPS}. This may be explained, in part,
by observational biases, since the existing IACT arrays have
relatively limited fields of view (up to 5$^{\circ}$). The new results
from HAWC, taking advantage of the larger field of view of this air
shower array, show that almost half of the HAWC-detected Galactic
sources have an angular size larger than 0.7$^{\circ}$ \cite{HAWC}.
\CTA{}, with an angular resolution as low as 0.05$^{\circ}$ at 1\,TeV
\cite{Gernot} and a maximum FoV of 10$^{\circ}$ (8$^{\circ}$) in the
southern (northern) array is a perfect instrument to perform
morphological studies of extended sources and to resolve structures
down to arcminute scales. We show the capabilities of \CTA{} for
morphological studies by looking at the specific case of the bright
young SNR, RX\,J1713.7-3946 \cite{j1713}; one of the key targets for
solving the problem of the CR origin. We created a template for our
simulations by combining a leptonic component traced by the XMM X-ray
image, and a hadronic component based on CO and HI observations
\cite{j1713cta}.  Figure \ref{fig:rx} compares the latest results
obtained by \hess{} with those expected for 50\,hr of observations
with \CTA{}, in the two extreme cases of leptonic and hadronic dominant
gamma-ray emission. As quantitatively proven in \cite{j1713cta},
\CTA{} will be able to distinguish between the two scenarios.

\section{Potential for Spectral Studies}
\begin{figure}
\begin{center}
\includegraphics[scale=0.45]{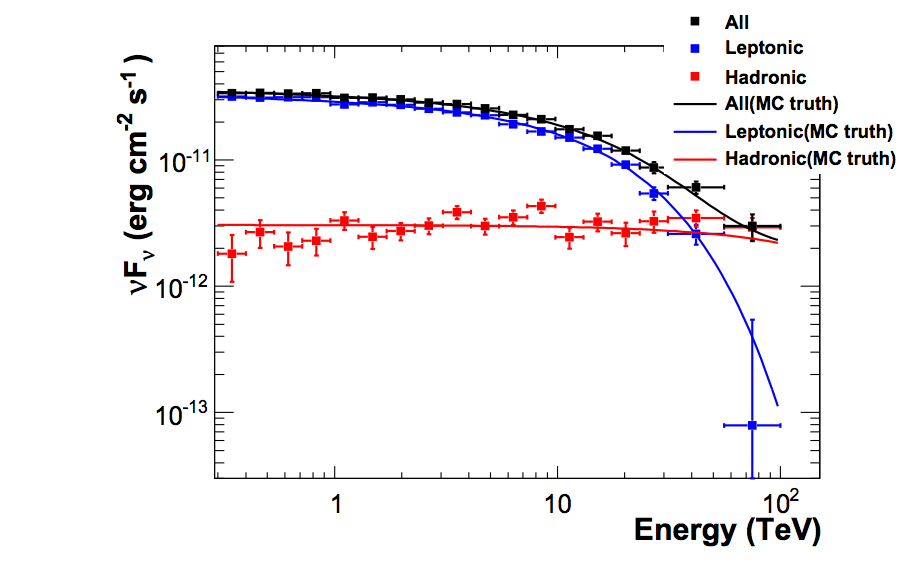}
\includegraphics[scale=0.45]{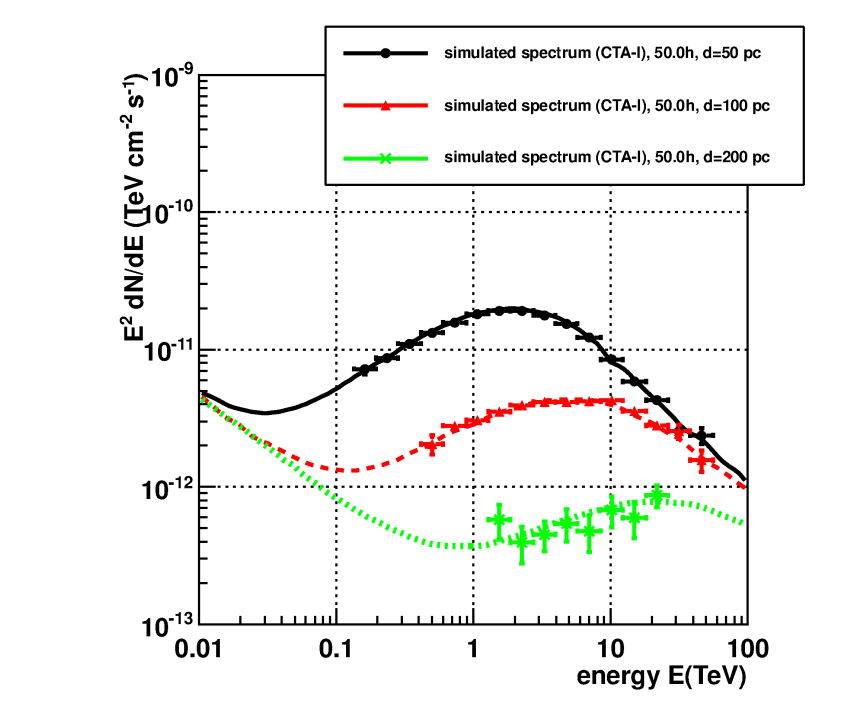}
\caption{\emph{On the left:} Spectral energy distribution of RX\,J1713.7-39. Blue and red
  squares are the spectral points for the leptonic and hadronic
  templates obtained for 50\,hr of observation. Black points are the
  total flux, the sum of the two components with error bars obtained by
  summing the two errors in quadrature. No systematic uncertainties
  are taken into account. Taken from \cite{j1713cta}.
\emph{On the right:} Simulated spectral energy distribution of a molecular cloud
  illuminated by a nearby SNR obtained with 50\,hr of
  observations. Different colours refer to different distances between
the cloud and the SNR. The molecular cloud is assumed to have a 10$^5$
M$_{\odot}$ and lie at 1\,kpc, whereas the SNR is 2000\,yr old. Taken
from \cite{Acero}. }
\label{fig:rx_sed}
\end{center}
\end{figure}

\CTA{}'s broadband coverage, together with its excellent energy
resolution (0.04\% at 10\,TeV \cite{Konrad}), will allow to detect
previously unmeasurable spectral features in Galactic sources. These
could include extra spectral components, such in the case of
RX\,J1713.7-3946. In the leptonic-dominant scenario of the emission
from this young SNR, CTA may detect a flatter hadronic component
emerging at very high energies, above 50\,TeV (see Figure
\ref{fig:rx_sed}), if the parent CR population is accelerated to more
than 100\,TeV. According to \cite{j1713cta}, such a detection, at
3$\sigma$ level, would require 50\,hr of observation by the southern
CTA{} array. Another example is given by the gamma-ray spectra of
molecular clouds illuminated by CRs from a nearby accelerator. In this
case, a characteristic V-shape may be imprinted on the spectral energy
distribution, due to the superposition of the gamma-ray emission from
background CRs with that from the nearby source \cite{Gabici,Acero}.
%{\bf not sure if to use the Crab or some other source. To be
 % discussed --> Emma does not want me to present the Crab results. I
 % will look for some others}. 
%On the other hand, we also studied \CTA{} capabilities to resolve the
%different cutoff models, i.e. sub-exponential and super-exponential,
%motivated by the interplay between acceleration and energy loss rates
%\cite{Kelner,Romoli}. We perform this study on the brightest known
%source, the Crab nebula. The statistical uncertainty after 5\,hr of
%observation by CTA-North (at small zenith angles) are small enough to
%distinguish model in which the cutoff exponent differs by more than
%0.66 (without accounting for systematic uncertainties). 

%\begin{figure}
%\begin{center}
%\includegraphics[scale=0.45]{rx_sed.png}
%\caption{Spectral energy distribution of the Crab Nebula}
%\label{fig:rx}
%\end{center}
%\end{figure}

\section{Conclusions}

The \CTA{} Galactic plane survey will lead to a significant
improvement in the understanding of our Galaxy at TeV energies, and
will allow to perform spectral and morphological studies with
unprecedented precision. \CTA{} can resolve spatial sub-structures on
arcminute scales, as well as reconstruct previously unmeasurable
spectral features. The combination of these new capabilities will be
particularly powerful, in that it provides the opportunity for
spatially resolved spectroscopy in the VHE gamma-ray regime. This is a
new window in TeV astronomy, which has previously required observation
exposures of typically hundreds of hours. In addition to superbly
detailed studies of specific, relatively bright objects, the GPS will
provide a new catalogue of VHE sources which will increase the source
count by a factor 3-9, and will very likely reveal new classes of
Galactic gamma-ray emitters.

\section*{Acknowledgments}
The CTA Consortium gratefully acknowledges financial support from the
following agencies and organizations listed here:
{\emph http://www.cta-observatory.org/consortium\_acknowledgments}

\end{document}